\documentclass{article}

\usepackage{arxiv}

\usepackage[utf8]{inputenc} 
\usepackage[T1]{fontenc}    
\usepackage{hyperref}       
\usepackage{url}            
\usepackage{booktabs}       
\usepackage{amsfonts}       
\usepackage{nicefrac}       
\usepackage{microtype}      
\usepackage{amsmath}
\usepackage{float}
\usepackage{varioref}

\usepackage{graphicx}
\newcommand{\Mathematica}{\textit{Mathematica\textsuperscript{\resizebox{!}{0.8ex}{\textregistered}}}}

\title{Scaling Index $\alpha = \frac{1}{2}$ In Turbulent Area Law}

\author{
  Alexander Migdal \\
  Fresnel Research LLC
  } 
\begin{document}
\maketitle

\begin{abstract}
We analyze the Minimal Area solution to the Loop Equations in turbulence \cite{M93}. As it follows from the new derivation in the recent paper  \cite{M19}, the vorticity is represented as a normal vector to the minimal surface not just at the edge, like it was assumed before, but all over the surface. As it was pointed in that paper, the self-consistency relation for mean vorticity leads to $\alpha=\frac{1}{2}$, however the similar conditions for product of two and more vorticities cannot be satisfied without extra terms, which were left undetermined in that paper. In this paper we find these missing terms -- they are delta functions at coinciding points which must be taken into considerations in surface integrals. We compare this value of $\alpha$ with new measurements of the same team which confirmed the area law \cite{S19} and we find that asymptotic formula $\lambda(p) \approx 2 \alpha p + \beta \ln p$, with  $\alpha =0.49 \pm 0.02, \beta= 0.92 \pm 0.01 $, fits all data at $p=3,...10$ within error bars.
\end{abstract}


\section{Introduction}
Let us summarize main equations of the loop dynamics as viewed from 21st Century.
The basic variable in the Loop Equations a circulation around closed loop in coordinate space

\begin{equation}\label{Gamma}
    \Gamma = \oint_C \vec v d\vec r 
\end{equation}

The PDF  for  velocity circulation as a functional of the loop 
\begin{equation}
    P\left ( C,\Gamma\right) =\left < \delta\left(\Gamma - \oint_C \vec v d\vec r\right)\right>
\end{equation}

with brackets \begin{math}< > \end{math} corresponding to time average or average over random forces, was shown to satisfy certain functional equation (loop equation). 

\begin{equation}\label{LoopEq}
\frac{\partial}{\partial \Gamma} \frac{\partial}{\partial t}  P\left ( C,\Gamma\right) 
=\oint_C d r_i \int d^3\rho\frac{ \rho_j }{4\pi|\vec \rho|^3}\frac{\delta^2 P(C,\Gamma)}{\delta \sigma_k(r) \delta \sigma_l(r + \rho)}\left(\delta_{i j}\delta_{k l} - \delta_{j k}\delta_{i l}\right)
\end{equation}

The  area derivative is defined using the difference between $P(C+\delta C,\Gamma)- P(C,\Gamma)$ where an infinitesimal loop $\delta C$ around the 3d point $r$ is added as an extra connected component of $C$. In other words, let us assume that the loop $C$ consists of an arbitrary number of connected components $C = \sum C_k$. 
We just add one more infinitesimal loop at some point away from all $C_k$. 
In virtue of the Stokes theorem, the difference comes from the circulation $\oint_{\delta C} \vec v d\vec r$ which reduces to vorticity at $r$
\begin{equation}
  P(C+\delta C,\Gamma)-P(C,\Gamma) =  d\sigma_i(r)    \left <\omega_i(r) \delta'\left(\Gamma - \oint_C \vec v d\vec r\right)\right>
\end{equation}
where
\begin{equation}
    d\sigma_k(r) = \oint_{\delta C} e_{i j k} r_i d r_j
\end{equation}
is an infinitesimal vector area element inside $\delta C$.
In general, for the Stokes type functional, by definition:
\begin{equation}
    U[C+\delta C]-U[C] = d\sigma_i(r) \frac{\delta U[C]}{\delta \sigma_i(r)}
\end{equation}
The Stokes condition $\oint_{\delta S} d \sigma_i \omega_i =0$ for  any closed surface $\delta S$ translates into
\begin{equation}\label{Stokes}
   \oint_{\delta S} d \sigma_i \frac{\delta U[C]}{\delta \sigma_i(r)} =0
\end{equation}

\begin{figure}[p]
    \centering
    \includegraphics{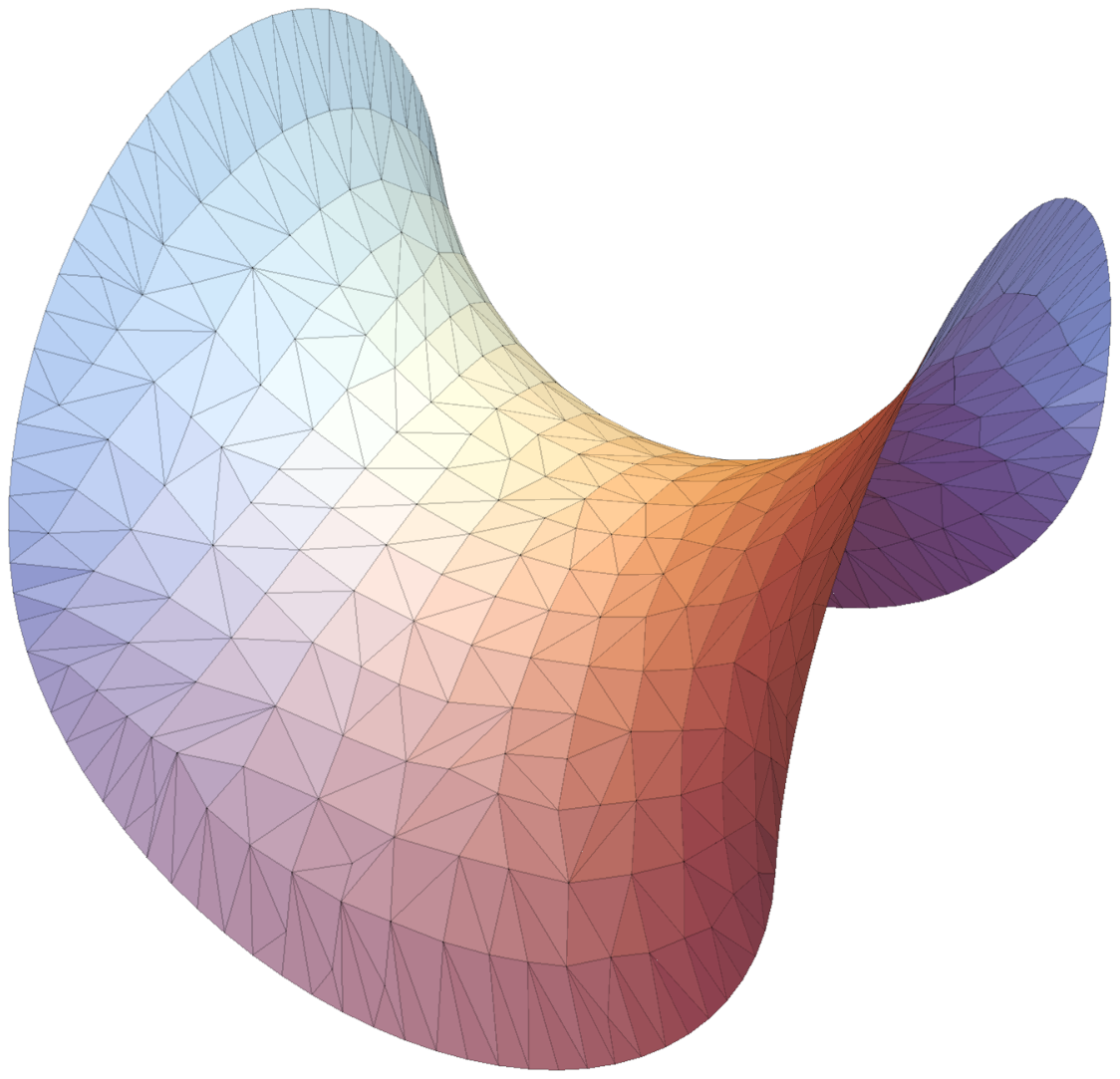}
    \caption{Minimal Surface, topologically equivalent to a disk (sphere with one hole), bounded by curved loop $C$.}
    \label{fig:OneLoop}
\end{figure}

The fixed point of the chain of the loop equations (\ref{LoopEq}) was shown to have solutions corresponding to two known distributions : Gibbs distribution and (trivial) global random rotation distribution. In addition, we found the third, nontrivial solution which is an arbitrary function of minimal area \begin{math} A_C \end{math} bounded by $C$.
\begin{equation}
    P(C,\Gamma) = F\left(A_C,\Gamma\right)
\end{equation}
The Minimal Area can be reduced to the Stokes functional by the following regularization
\begin{equation}\label{Regularized}
  A_C =\min_{S_C} \int_{S_C} d \sigma_{i}(r_1) \int_{S_C} d \sigma_{j}(r_2) \delta_{i j} \Delta(r_1-r_2)
\end{equation}
with 
\begin{equation}
\Delta(r) = \frac{1}{r_0^2} \exp\left(-\pi \frac{r^2}{r_0^2}\right); r_0 \rightarrow 0
\end{equation}
representing two dimensional delta function, and integration goes over minimized surface $S_C$ (see Fig.\ref{fig:OneLoop}, created with \Mathematica\ \cite{MinSurfaceX}).

\begin{figure}[p]
    \centering
    \includegraphics{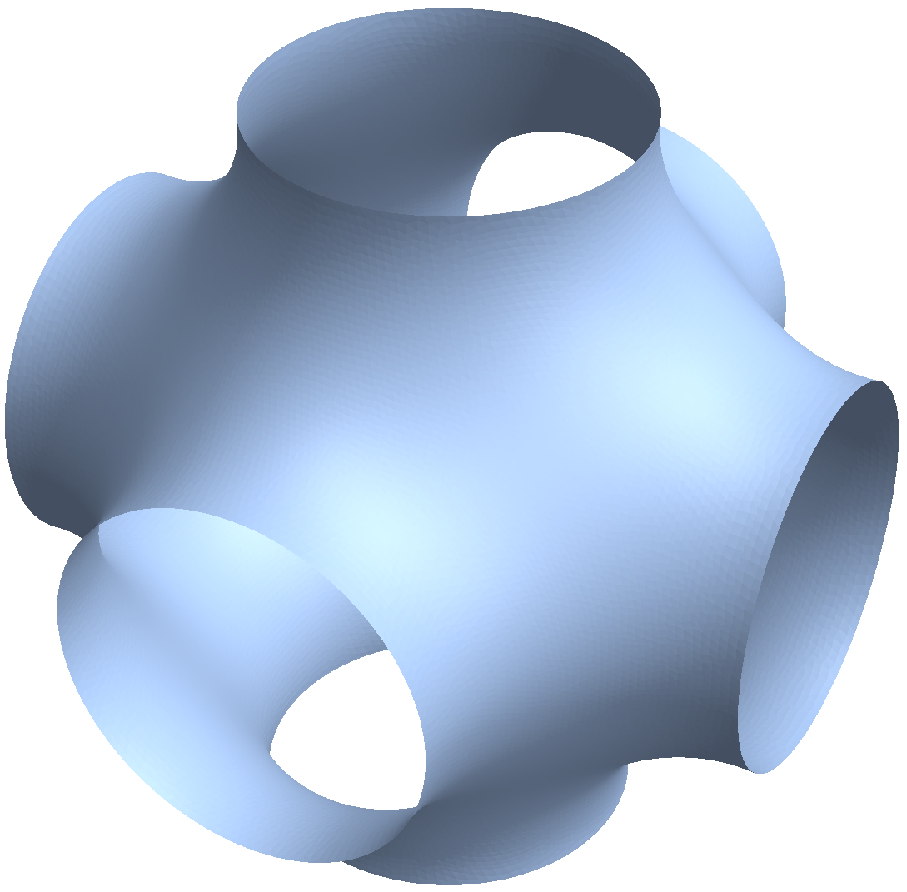}
    \caption{Minimal Surface, topologically equivalent to a sphere with $N=6$ holes, bounded by loop $C= \sum_{k=1}^6 C_k$.}
    \label{fig:MultiLoop}
\end{figure}

In general case the loop $C$ consist of $N$ closed pieces $C = \sum_{k=1}^N C_k$ and the surface $S_C$ must connect them all, so that it is topologically equivalent to a sphere with $N$ holes and no handles (see Fig.\ref{fig:MultiLoop} created with \Mathematica \cite{MinSurfaceX}).
In case some pieces are far away from others, the minimal surface would make thin tubes reaching from one closed loop $C_k$ to another via some central hub where all the tubes grow out of the sphere.
In our case we need only two extra little loops both close to the initial contour $C$ but for completeness we must assume there is an arbitrary number of closed pieces in $C$.

In real world this $r_0$ would be the viscous scale $\left(\frac{\nu^3}{\cal E}\right)^{\nicefrac{1}{4}}$.
This is a positive definite functional of the surface as one can easily verify using spectral representation:
\begin{equation}
    \int_{S_C} d \sigma_{i}(r_1) \int_{S_C} d \sigma_{j}(r_2) \delta_{i j} \Delta(r_1-r_2) \propto \int d^3k \exp\left(-\frac{k^2r_0^2}{4\pi}\right) \left| \int_{S_C} d \sigma_{i}(r)e^{i k r} \right|^2
\end{equation}
In the limit $r_0 \rightarrow 0$ this definition reduces to the ordinary area:
\begin{equation}
  A_C \rightarrow \min_{S_C} \int_{S_C} d^2 \xi \sqrt{g} 
\end{equation}

The Stokes condition (\ref{Stokes}) is satisfied in virtue of extremum condition. When the surface changes into $S'$, so that the linear variation reduces to the integral (\ref{Stokes}) with $\delta S = S'-S$ being the infinitesimal closed surface between $S'$ and $S$. This linear variation must vanish by definition of the minimal surface,  for regularized area as well as for its local limit.

The area derivative of the Minimal Area in regularized form, then, as before, reduces to elimination of one integration
\begin{equation}\label{vorticity}
  \frac{\delta A_C}{\delta \sigma_i(r)} = 2 \int_{S_C} d \sigma_{i}(\rho) \Delta(r-\rho) \rightarrow 2 n_i(\Tilde{r})\exp\left(-\pi \frac{r_{\perp}^2}{r_0^2}\right)
\end{equation}
Where $n_i(\Tilde{r}) $ is the local normal vector to the minimal surface at the nearest surface point $\Tilde{r}$ to the 3d point $r$, and $r_{\perp}$ is the component normal to the surface at $\Tilde{r}$. With this regularization area derivative is defined everywhere in space but it exponentially decreases away from the surface. Exactly at the surface it reduces to twice the unit normal vector\footnote{As for Stokes condition $\partial_i \frac{\delta A_C}{\delta \sigma_i(r)} =0$ one can readily check in a local coordinate frame where $\vec r = (x,y,z) $ and the surface equation is $ z = \frac{1}{2} \left(k_1 x^2 + k_2 y^2\right)$, that at $r_0 \rightarrow 0$ the Stokes condition reduces to $\int_{-\infty}^{\infty}d x \int_{-\infty}^{\infty} d y \partial_z \Delta (\vec r) \propto (k_1 + k_2)=0 $ which is the well known equation of vanishing mean curvature at a minimal surface.}.

\begin{figure}[p]
    \centering
    \includegraphics{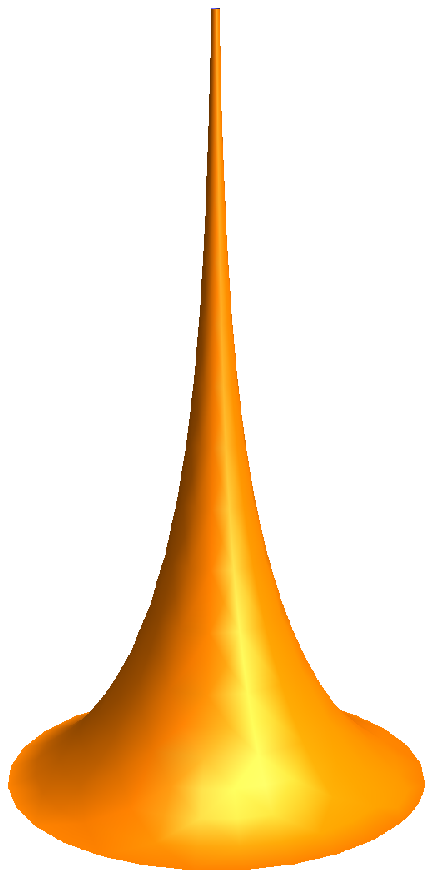}
    \caption{Minimal Surface, stretched to reach a remote point (infinitesimal loop) }
    \label{fig:Peak}
\end{figure}

Should we go to the limit $r_0 \rightarrow 0$ first we would have to consider the minimal surface connecting the original loop $C$ and infinitesimal loop $\delta C$. Such minimal surface would have a thin tube connecting the point $r$ to  $\Tilde{r}$ at the original minimal surface along its local normal $\vec n$ (see Fig.\ref{fig:Peak}).
We are not going to investigate this complex problem here -- with the regularized area we have explicit formula, and we need this formula only at the boundary in leading log approximation (see \cite{M19}).

As we argued in old paper \cite{M93} we expect scale invariant solutions, depending of $\gamma =\Gamma A_C^{-\alpha}$ in our scale invariant equations, with some critical index $\alpha$, yet to be determined.
Let us stress again, that the Kolmogorov value of the scaling index $\alpha_{K} = \frac{2}{3}$ does \textbf{not} follow from the loop equations, this is an additional assumption, based on dimensional counting and Kolmogorov anomaly \cite{M93} for the third moment of velocity. As it was stressed in that paper, the Kolmogorov anomaly poses no restrictions on the vorticity correlations, and cannot therefore be used to determine our scaling index. 

The Area law is expected to be an asymptotic solution at large enough circulations and areas where PDF is small. Such PDF tails are  usually interpreted as instantons or classical solutions in some variables \cite{FLM}. In our variables this is the minimal area as a functional of the its boundary loops  $C = \sum_k C_k$ . \footnote{This observation suggests that there may be some functional integral for PDF where the classical action would be the surface area and the sum over histories goes over all surfaces bounded by $C$. In short, that would be a string theory of some sort.
Nobody managed so far to make sense out of string theory in 3 dimensions -- it looks like in any dimension less than $26$ the surface degenerates into a branched polymer. Our arguments in the loop equations do not imply such string theory analogy. There are significant differences between the Loop equation in fluid dynamics and that in QCD where we know that the area law holds asymptotically (there are singularities at self-intersections of the loop in QCD). But even in QCD nobody managed to present the solution to the loop equation as some kind of a string theory.}

This Universal Area Law  was confirmed in numerical experiments \cite{S19} with Reynolds up to $10^4$ with very high accuracy over whole inertial range of circulations and areas with PDF from 1 down to $10^{-8}$.  See \cite{M19} for analysis of numerical results and their correspondence with Area Law.

Let us present more conventional physical interpretation of this area law.

Let us consider vorticity field (\ref{vorticity}) generated by a minimal surface with thickness $r_0$. It rapidly decreases outside the surface and equals twice the normal vector at the surface.
Corresponding velocity field will be defined everywhere in space by the integral
\begin{equation}
    v_i(r) \propto  e_{i j k}\int d^3 r'\frac{r'_j -r_j}{|\vec r - \vec r'|^3} n_k(\Tilde{r}') \Delta(r-r')
\end{equation}

In particular, directly at the surface
\begin{equation}
    v_i(r) \propto r_0 e_{i j k}\int_{S_C} d^2 r'\frac{r'_j -r_j}{|\vec r - \vec r'|^3} n_k(r')
\end{equation}
If $r$ approaches the edge, this integral logarithmically diverges (we use the frame where surface normal is directed to $z$, the local tangent to $C$ goes along $x$ and inside direction into the surface goes along $y$ and $\rho =x,y$ are local coordinates on the surface near its edge.
\begin{equation}
    v_a(C - \epsilon) \propto r_0 e_{a b }\int_{-\infty}^{\infty} d x \int_0^\infty d y \rho_b \left((y+\epsilon)^2 + x^2\right)^{-\frac{3}{2}} \propto r_0 \delta_{a 1 } \int_0^\infty d y \frac{y}{(y+\epsilon)^2} \propto r_0 \delta_{a 1 }\ln \epsilon
\end{equation}
Now, as is well known (Kelvin theorem, see also \cite{M19})
\begin{equation}
    \partial_t \Gamma \propto e_{i j k} \int_C d r_i v_j \omega_k 
\end{equation}
In our coordinate frame, using the fact that $\vec \omega \propto \vec n$ is directed along $z$
\begin{equation}
    \partial_t \Gamma \propto  \int_C d x v_2   
\end{equation}
which vanished as $v_2=0$. 
So, the circulation would  be conserved for this particular vorticity distributed in infinitesimal (i.e viscous) layer around minimal surface.

We have shown that this is indeed a stationary PDF for the circulation. However, velocity field is far from being conserved. The nonlinear term $ v_k \partial_k v_i$ does not vanish anywhere in space, including the minimal surface with this vorticity field. Only after cancellation of $\nabla \left(\frac{v^2}{2} + p\right)$ in the integral over closed loop for $\partial_t \Gamma $ do we get simple local term $e_{i j k} d r_i v_j \omega_k$ which, as we proved, cancels in a leading logarithmic approximation.

\section{Equation for Scaling Index}

Let us start without assumption of scaling law, with weaker assumption of some unknown function of the minimal area as the scale of $\Gamma$
\begin{equation}\label{AutoModel}
     P(C,\Gamma)\rightarrow G(\ln A_C)\Pi\left(\Gamma G(\ln A_C)\right) 
\end{equation}
The factor of $G(\ln A_C)$ in front of scaling function $\Pi(\gamma)$ follows from the fact that $\Gamma P(C,\Gamma)$ must be scale invariant, regardless how effective scale $G$ depends on $\ln A_C$.

Let us derive self-consistency equation for $G(\ln A)$.

On one hand:
\begin{align}
    \left< \int_{S_C} d\vec \sigma \vec \omega \delta'\left(\Gamma - \int_{S_C} d\vec \sigma \omega\right) \right>
    &= \partial_\Gamma\left(\left< \int_{S_C} d\vec \sigma \vec \omega \delta\left(\Gamma - \int_{S_C} d\vec \sigma \omega\right)\right> \right)\\ 
    &=\partial_\Gamma \left(\Gamma\left< \delta\left(\Gamma - \int_{S_C} d\vec \sigma \omega\right)\right>\right)\\
    &=\partial_\Gamma \left(\Gamma G(\ln A_C) \Pi\left(\Gamma G(\ln A_C)\right) \right) \label{One}
\end{align}
On another hand:
\begin{align}
    \left< \int_{S_C} d\vec \sigma\vec \omega \delta'\left(\Gamma - \int_{S_C} d\vec \sigma \omega\right) \right>
     &= -\left<\int_{S_C} d\vec \sigma \frac{\delta}{\delta \vec \sigma} \delta\left(\Gamma - \int_{S_C} d\vec \sigma \omega\right)\right>\\
    &= -\int_{S_C} d\vec \sigma \frac{\delta}{\delta \vec \sigma}\left< \delta\left(\Gamma - \int_{S_C} d\vec \sigma \omega\right)\right>\\
    &= -\int_{S_C} d\vec \sigma \frac{\delta}{\delta \vec \sigma}\left( G(\ln A_C) \Pi\left(\Gamma G(\ln A_C)\right) \right)\label{Second}
\end{align}
Recall that at the minimal surface
\begin{align}
    \frac{\delta A_C}{\delta \vec \sigma} &= 2 \vec n\\
    \int_{S_C} d\vec \sigma \vec n &= A_C
\end{align}

and we find in (\ref{One}):
\begin{equation}
    G \left( \Pi(\Gamma G) + \Gamma G \Pi'(\Gamma G)\right)
\end{equation}
and in (\ref{Second}):
\begin{equation}
   -2 \frac{\partial}{\partial \ln A_C} \left( G(\ln A_C) \Pi\left(\Gamma G(\ln A_C)\right) \right)=
   -2G'\left( \Pi(\Gamma G) + \Gamma G \Pi'(\Gamma G)\right)
\end{equation}
Comparing these two expressions we find the differential equation
\begin{equation}
    G' = -\frac{1}{2} G
\end{equation}
with solution
\begin{equation}
    G \propto A_C^{-\frac{1}{2}}
\end{equation}
Note that there is no restriction on the PDF scaling function $\Pi(\gamma)$.

Note also that as $\Gamma$ changes sign on time reversal, we expect the dissipation reflect itself in asymmetry of PDF. Indeed, as measured in \cite{S19} the left tail of the PDF decreases with twice larger slope in stretched exponential decay, compared to the right tail. As a consequence the odd moments $\left< \Gamma^p \right>$ are present even at large $p$ where right saddle point dominates, and the left saddle point provides exponentially small correction (see discussion below).

\section{Higher Correlations}
Now let us check that this remarkable solution is compatible with the higher correlations.

Let us recall  the results  \cite{M19}:
\begin{equation}\label{MultiNormal}
    \left < \vec \omega_1\dots \vec\omega_k \delta\left(\Gamma - \oint_C \vec v d\vec r\right)\right >  = \vec n_1 \dots \vec n_k A_C^{-\alpha - k(1-\alpha)}\Omega_k\left(\Gamma A_C^{-\alpha}\right)
\end{equation}
The scaling functions $\Omega_k(\gamma)$ with $\Omega_0(\gamma) = \Pi(\gamma)$ being scaling PDF, satisfy recurrent equations :
\begin{equation}\label{Recurrent}
    \Omega_{k+1}\left(\gamma\right)= 2 \alpha \gamma\Omega_{k}\left(\gamma\right)  -2(1-\alpha)k\int_\gamma^{\pm \infty}\Omega_{k}(y) \,d y
\end{equation}
 \begin{equation}
   \left < \delta\left(\Gamma - \oint_C \vec v d\vec r\right)\right> = A_C^{-\alpha}\Pi\left(\frac{\Gamma}{A_C^{\alpha}}\right)
 \end{equation}
 \begin{equation} 
     \left < \vec \omega \delta\left(\Gamma - \oint_C \vec v d\vec r\right)\right> =2\alpha\vec n \frac{\Gamma}{A_C} \Pi\left(\frac{\Gamma}{A_C^{\alpha}}\right)
 \end{equation}
 
 From original derivation using the area derivative it follows that this equation holds on \textbf{the whole surface} as well as at its edge $C$.
 Therefore, as pointed out in \cite{M19} we can integrate this equation over the surface:
 \begin{equation}
   \left <  \int_{S_C} d \vec \sigma(r) \vec \omega(r) \delta\left(\Gamma - \oint_C \vec v d\vec r\right)\right> =2\alpha\int_{S_C} d \vec \sigma(r)\vec n(r) \frac{\Gamma}{A_C} \Pi\left(\frac{\Gamma}{A_C^{\alpha}}\right)
 \end{equation}
 
 On the left side we obtain $\int_{S_C} d \vec \sigma(r) \vec \omega(r) = \Gamma$ in virtue of the $\delta$ function, and on the right side we get
 $\int_{S_C} d \vec \sigma(r)\vec n(r) = A_C$. As a result, after cancelling $\Gamma$ we obtain equation:
 \begin{equation}
     \Pi\left(\frac{\Gamma}{A_C^{\alpha}}\right) = 2 \alpha \Pi\left(\frac{\Gamma}{A_C^{\alpha}}\right)
 \end{equation}
from which we conclude that
\begin{equation}
    \alpha = \frac{1}{2}
\end{equation}
as we already found in the previous section.

With the next equation though, things are getting tricky, and this is where we got stuck in \cite{M19}:
 \begin{equation} 
     \left < \omega_1 \omega_2 \delta\left(\Gamma - \oint_C \vec v d\vec r\right)\right> = n_1  n_2  A_C^{-\frac{3}{2}}\Omega_2\left(\Gamma A_C^{-\frac{1}{2}}\right)
 \end{equation}
with 
\begin{equation}\label{Omega2}
    \Omega_2(\gamma) = \gamma^2\Pi(\gamma) -\int_{ \gamma}^{\pm\infty} \Pi(y)y \,d y 
\end{equation}
Integrating over the surface we get, as before, after going to scaling variables, and setting $\alpha=\frac{1}{2}$, on the left :
\begin{equation}
    \gamma^2\Pi(\gamma)
\end{equation}
and on the right
\begin{equation}
    \gamma^2\Pi(\gamma)  -\int_{ \gamma}^{\pm\infty} \Pi(y)y \,d y 
\end{equation}
Everything stops! The left part does not match the right part.

Here is what we forgot: the contact terms. The correlations of vorticity were obtained assuming the points do not coincide. In general, we can expect extra contact term:
\begin{equation} \label{ContactTerm}
     \left < \omega_1 \omega_2 \delta\left(\Gamma - \oint_C \vec v d\vec r\right)\right> = n_1  n_2  A_C^{-\frac{3}{2}}\Omega_2\left(\Gamma A_C^{-\frac{1}{2}}\right) + X \delta_{1 2} \delta^2(1-2)
 \end{equation}
where $X$ is to be determined, and $\delta_{1 2}$ is Kronecker delta for vector indexes and $\delta^2(1-2)$ is an invariant delta function on the surface. These contact terms display themselves only in the integral relations we have here, and do not change correlations at far away points (which was assumed in \cite{M19}).
With this term present we get perfect match if
\begin{equation}
    X = \int_{ \gamma}^{\pm\infty} \Pi(y)y \,d y 
\end{equation}
What could be the origin of such term? Let us consider the second derivative of (\ref{ContactTerm}) by $\Gamma$. On the left we find 
\begin{equation}
    \left < \omega_1 \omega_2 \delta''\left(\Gamma - \oint_C \vec v d\vec r\right)\right>=
    \frac{\delta^2}{\delta \sigma(1)\delta \sigma(2)}\left < \delta\left(\Gamma - \oint_C \vec v d\vec r\right)\right> 
    =\frac{\delta^2}{\delta \sigma(1)\delta \sigma(2)} A_C^{-\frac{1}{2}} \Pi\left(\frac{\Gamma}{A_C^{\frac{1}{2}}}\right)
\end{equation}
The contact term precisely of the form we need comes from the second functional derivative of $A_C$
\begin{equation}
    \frac{\delta^2 A_C}{\delta \sigma(1)\delta \sigma(2)}\partial_{A} A^{-\frac{1}{2}} \Pi\left(\Gamma A^{-\frac{1}{2}}\right) =  2\delta_{1 2} \delta^2(1-2)\partial_{A} A^{-\frac{1}{2}} \Pi\left(\Gamma A^{-\frac{1}{2}}\right)
\end{equation}

The rest of the terms were accounted in recurrent equations for vorticity expectation values in \cite{M19}, but this one was missed (or, better to say, ignored as we were assuming all points separate).
The derivative of $X$ term matches this one as
\begin{equation}
    \partial_\gamma^2\int_{ \gamma}^{\pm\infty} \Pi(y)y \,d y  = - \left(\Pi(\gamma)+ \gamma \Pi'(\gamma)\right)
\end{equation}
and
\begin{equation}
  2 \partial_{A} A^{-\frac{1}{2}} \Pi\left(\Gamma A^{-\frac{1}{2}}\right)\propto -\left(\Pi(\gamma)+ \gamma \Pi'(\gamma)\right)
\end{equation}
Here we dropped factors of $A_C$ as they are known to match by dimensional counting.
So, the contact terms from the second functional derivative precisely match missing terms in the self-consistency relation for the two point function.

Let us present the result for the next correlation functions with corrected coefficients at $\alpha= \frac{1}{2}$, which were created in \Mathematica\ by means of symbolic integration by parts\footnote{Courtesy of Arthur Migdal, 2d year, MIT.}:
\begin{align}
\Omega_{1}(\gamma) &= \gamma  \Pi (x)\\
\Omega_{2}(\gamma) &= \gamma ^2 \Pi (x)-\int_{\gamma }^{\pm \infty } y \Pi (y) \, d y\\
\Omega_{3}(\gamma) &= \gamma ^3 \Pi (x)-3 \int_{\gamma }^{\pm \infty } \gamma  y \Pi (y) \, d y\\
\Omega_{4}(\gamma) &= \gamma ^4 \Pi (x)+\frac{3}{2} \int_{\gamma }^{\pm \infty } y \left(y^2-5 \gamma ^2\right) \Pi (y) \, d y\\
\Omega_{5}(\gamma) &= \gamma ^5 \Pi (x)+\frac{5}{2} \int_{\gamma }^{\pm \infty } \gamma  y \left(3 y^2-7 \gamma ^2\right) \Pi (y) \, d y\\
\Omega_{6}(\gamma) &= \gamma ^6 \Pi (x)-\frac{15}{8} \int_{\gamma }^{\pm \infty } y \left(21 \gamma ^4+y^4-14 \gamma ^2 y^2\right) \Pi (y) \, d y\\
\Omega_{7}(\gamma) &= \gamma ^7 \Pi (x)-\frac{21}{8} \int_{\gamma }^{\pm \infty } \gamma  y \left(33 \gamma ^4+5 y^4-30 \gamma ^2 y^2\right) \Pi (y) \, d y\\
\Omega_{8}(\gamma) &= \gamma ^8 \Pi (x)+\frac{7}{16} \int_{\gamma }^{\pm \infty } y \left(-429 \gamma ^6+5 y^6-135 \gamma ^2 y^4+495 \gamma ^4 y^2\right) \Pi (y) \, d y\\
\Omega_{9}(\gamma) &= \gamma ^9 \Pi (x)+\frac{9}{16} \int_{\gamma }^{\pm \infty } \gamma  y \left(-715 \gamma ^6+35 y^6-385 \gamma ^2 y^4+1001 \gamma ^4 y^2\right) \Pi (y) \, d y\\
\Omega_{10}(\gamma) &= \gamma ^{10} \Pi (x)-\frac{45}{128} \int_{\gamma }^{\pm \infty } y \left(2431 \gamma ^8+7 y^8-308 \gamma ^2 y^6+2002 \gamma ^4 y^4-4004 \gamma ^6 y^2\right) \Pi (y) \, d y\\
\Omega_{11}(\gamma) &= \gamma ^{11} \Pi (x)-\frac{55}{128} \int_{\gamma }^{\pm \infty } \gamma  y \left(4199 \gamma ^8+63 y^8-1092 \gamma ^2 y^6+4914 \gamma ^4 y^4-7956 \gamma ^6 y^2\right) \Pi (y) \, d y
\end{align}
\section{Discussion}
The main result of this work is exact computation of the critical index from self-consistency of the Minimal Area solution of the Loop Equations. With correct recurrent equations (\ref{Recurrent}) the contact terms needed for consistency of surface integrals of the vorticity correlation functions, naturally arise from second functional derivatives of the regularized Area (\ref{Regularized}).

It looks like the ends start meeting in this exotic solution, which initially raised so much confusion 26 years ago. But does it meet the numerical experiment? 

Initially, the authors of that experiment \cite{S19} interpreted their results for the log-log derivative 
\begin{equation}\label{lambda_p}
    \lambda(p) = \frac{d \log \left<|\Gamma|^p\right>}{d \log \sqrt{A_C}}
\end{equation}
as some bi-fractal model, switching  from K41 value $\lambda(p) =\frac{4p}{3}$ at $ p < 4$ to another slope $\lambda(p) \approx 1.16 p$ at $4 \leq p \leq 10$. The linear fit of their data is not perfect; one can see that the slope $\frac{\lambda(p)}{p}$, is never a constant -- it is greater than $\frac{4}{3}$ at small p and steadily decreases.

According to our theory here, it should reach $\lambda(p) \rightarrow p$.
The finite moments are not expected to obey the Area law -- this is an opposite limit of large $\Gamma$ which corresponds to the saddle point $\Gamma_0$ in the moments integral after switching to $\ln \Gamma$ as integration variable 
\begin{equation}\label{MomentIntegral}
    \left<|\Gamma|^p\right> = \int_{0}^{\infty} d \Gamma \Gamma^p P(\Gamma,C) \propto \Gamma_0^{p+1} P(\Gamma_0,C)\left(-\left(\Gamma_0 \partial_{\Gamma_0}\right)^2 \ln P(\Gamma_0,C)\right)^{-\frac{1}{2}} (1+ \dots);
\end{equation}
\begin{equation}\label{Saddle}
    p + 1 + \Gamma_0 \partial_{\Gamma_0}\ln \left(P(\Gamma_0,C)\right) =0
\end{equation}
Here $\dots$ stand for higher corrections to the saddle point integration, related to higher derivatives $\left(\Gamma_0 \partial_{\Gamma_0}\right)^n \ln P(\Gamma_0,C), n>2 $ in Taylor expansion near the saddle point. For power-like $\ln P(\Gamma,C)$ these corrections go in inverse powers of $p$.

Clearly the largest of two possible saddle points at positive and negative $\Gamma$ determines the asymptotic  behavior of the moments. Another saddle point provides exponential corrections at large $p$. We expect the positive saddle point to dominate,  as it decreases slower in numerical experiments \cite{S19}, so we  only use the $\Gamma >0$ integral.

This saddle point at large $p$ would approach the region of large $\Gamma$ where we expect Area law to hold.
Technically it is simpler to treat $\Gamma_0$ as  an independent large variable and the saddle point equation (\ref{Saddle}) as  parametric equation for $p(\Gamma_0)$. In that case all the terms can be explicitly computed to any order in the WKB expansion around the saddle point.

In practice such saddle point approximations may be very accurate even for small $p$ as the Stirling formula for the Gamma function shows. 
Apparently, this numerical luck explains validity of the Area Law in a whole inertial range and even small moments.

Let us assume the following correction 
\begin{equation}
    \lambda(p) \rightarrow 2\alpha p + \beta \ln p
\end{equation}
where the coefficient $\beta$ is not universal and can depend on the shape of the loop  as well as the area.\footnote{This $\beta$ term corresponds to some pre-exponential power of $p^{\xi}$ in the moment $\left<\Gamma^p\right>$ with the index $\xi$ depending on $\ln A_C$ so that $\beta = 2 \partial_{\ln A} \xi$.}

\begin{figure}[p]
    \centering
    \includegraphics{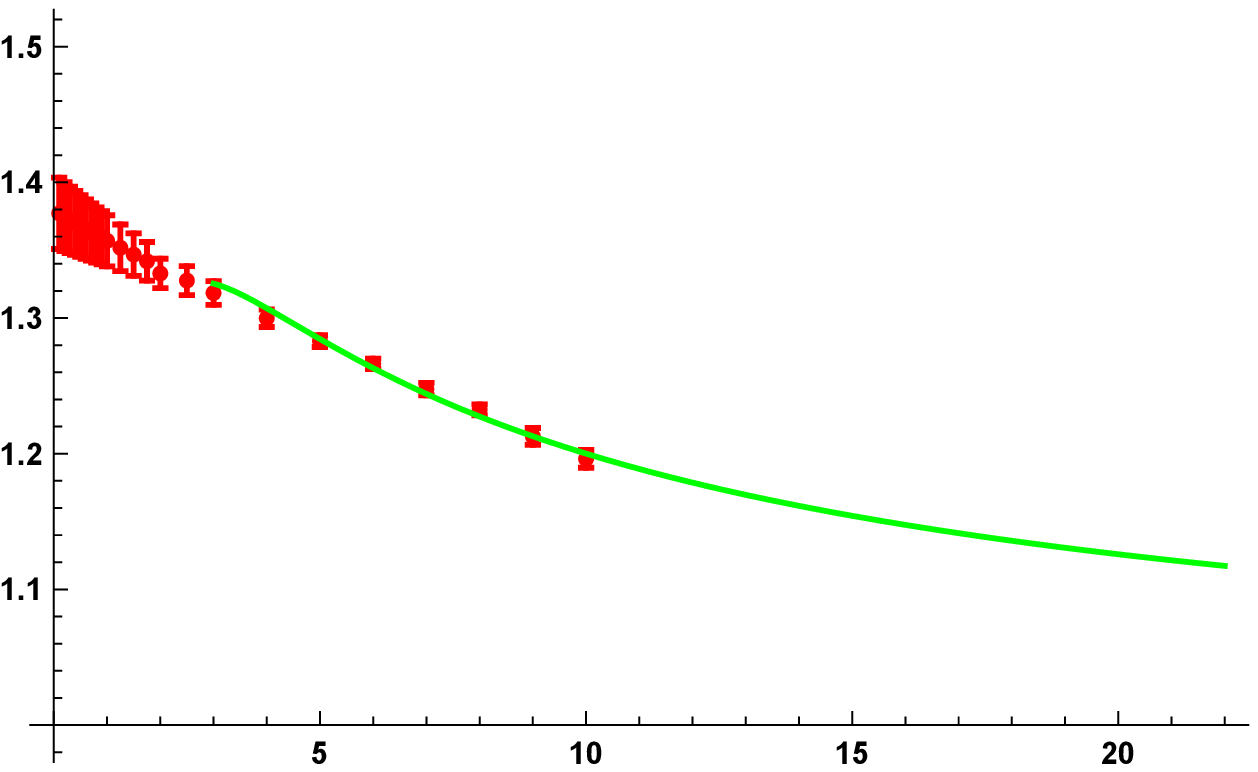}
    \caption{Effective index $\frac{\lambda(p)}{p}$ from \cite{S19} with our asymptotic fit (green): $\frac{\lambda(p)}{p} \approx  0.987832 + 0.921675 \frac{\log p}{p}$ }
    \label{fig:effective_index}
\end{figure}

The $p$ dependence can be compared with  experimental data\footnote{This data was shared with me by Kartik P. Iyer.} at maximal available Reynolds number $1300$. The following parameters fit the data at $p=3,...10$ with adjusted $R^2= 0.999991$ which means extremely well:
\begin{align}
    \alpha &=0.49 \pm 0.02\\
    \beta &= 0.92 \pm 0.01
\end{align}
We performed statistical analysis with \Mathematica\ .
Let us compare with experiment our asymptotic formula for the ratio $\frac{\lambda(p)}{p}$.  (see Fig. \ref{fig:effective_index}). 
As we can see, it the green curve passes within error bars through all the data points at $p=3,...10$.

For those of you who do not believe in fitting, here is the plot of pure experimental data with error bars for $\frac{\lambda(p)-p}{\ln p}$ in a wider range of $p$ including small $p$ where we do not expect it to be constant. 
\begin{figure}[p]
    \centering
    \includegraphics{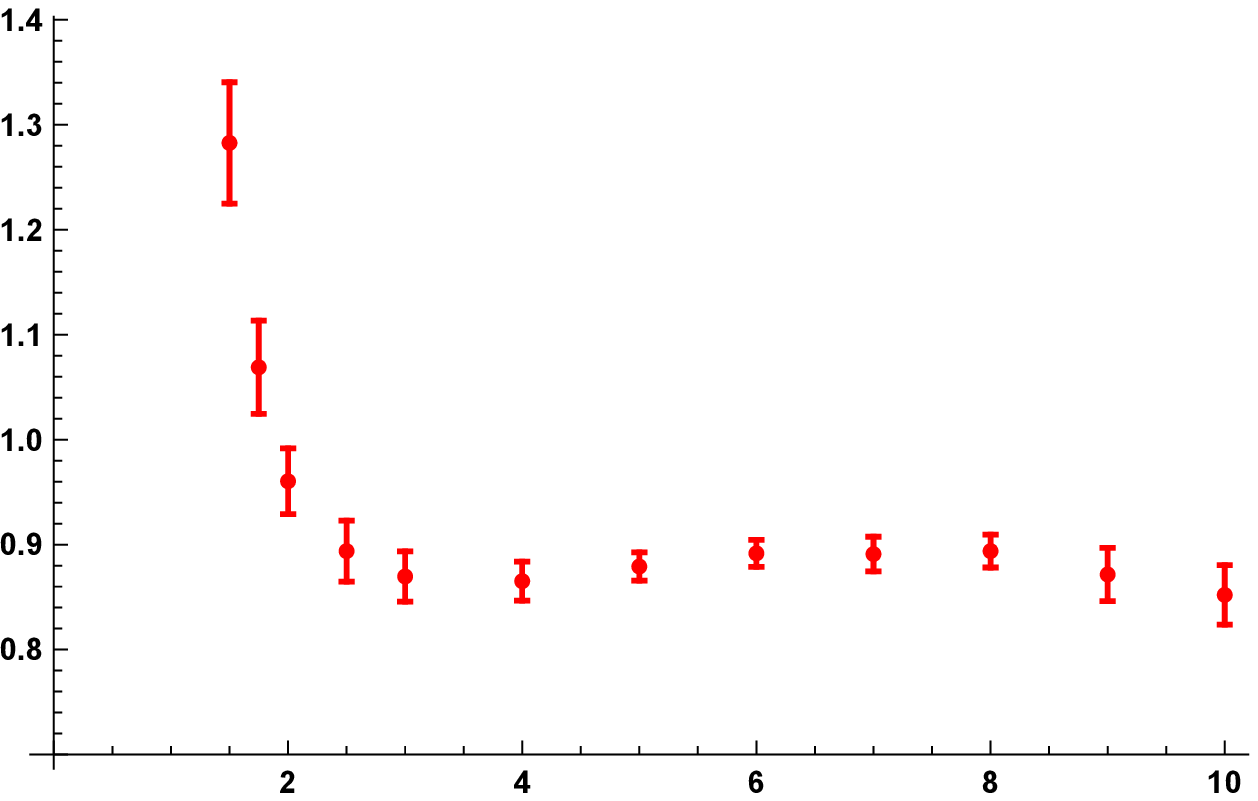}
    \caption{Experimental data for the ratio $\frac{\lambda(p)-p}{\ln p}$ from \cite{S19}. }
    \label{fig:is_it_log}
\end{figure}
Apparently it is consistent with assumption of a constant limit, just take a piece of paper and place its edge across the data points.

We take it as another validation of the Area Law, though more experiments at higher Reynolds number would be needed to verify asymptotic index $\alpha = \frac{1}{2}$ with two or more significant digits. The ratio $\frac{\lambda(p)}{p}$ is clearly falling up to $p=10$ and would keep falling after that as $\frac{\ln p}{p}$ according to our current fit. So, what happens with experiment for $p =11,...20$?.

\begin{figure}[p]
    \centering
    \includegraphics{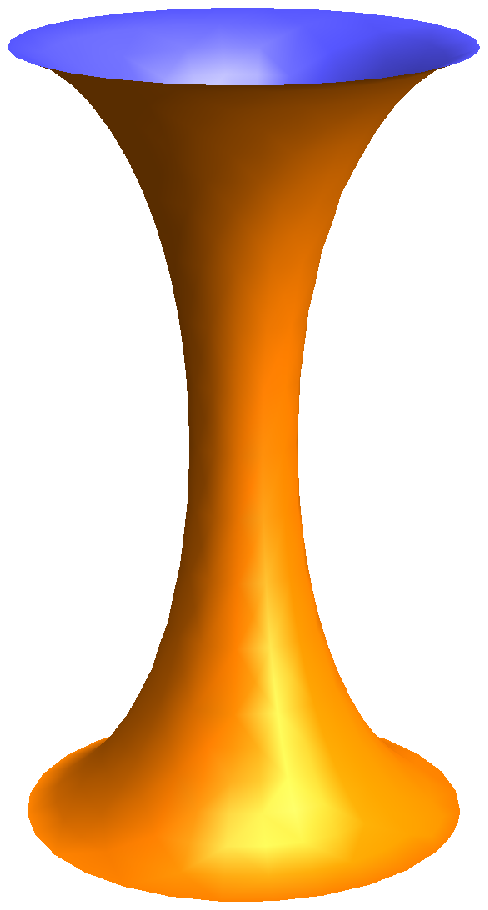}
    \caption{Tube-like minimal surface connecting two separated loops.}
    \label{fig:Tube}
\end{figure}

If we start believing the Minimal Surface we have to try and see how the correlation between two circulations $\Gamma_1, \Gamma_2$ around planar loops  in $x y$ plane depends upon the separation between these loops in the orthogonal direction $z$. (see Fig.\ref{fig:Tube}).

When the small loop starts inside the big one and moves in $z$ direction, the minimal surface would grow like a tower between these two loops. In the limit of small loop much less that the big one (see Fig.\ref{fig:Peak}) we must approach the vorticity correlation as a function of the normal distance to the minimal surface for the large square (see Fig. \ref{fig:Peak}). How does that correlation depends of $z$? 

\begin{figure}[p]
    \centering
    \includegraphics{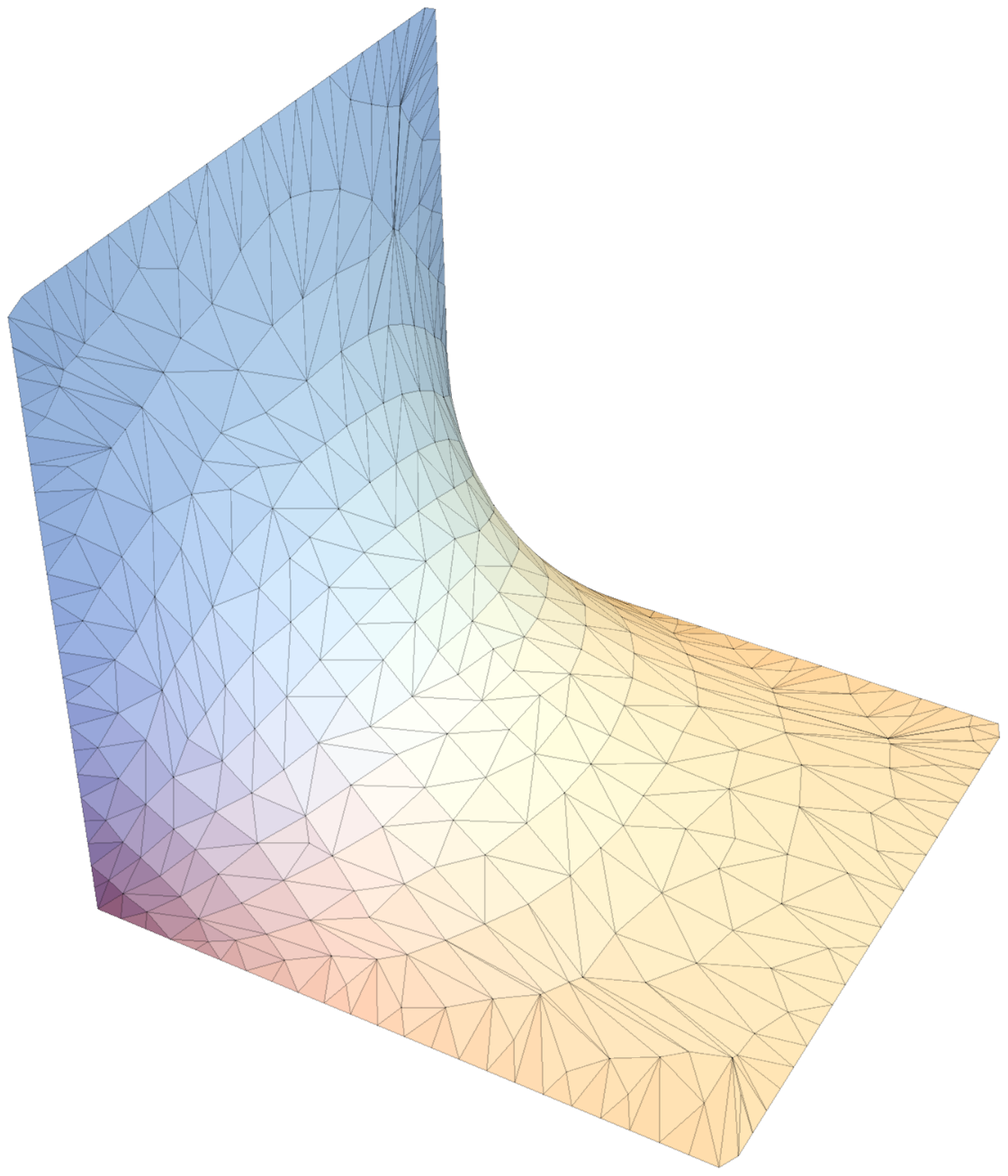}
    \caption{Minimal Surface bounded by soccer gates.}
    \label{fig:SoccerGates}
\end{figure}

Also, the "soccer gate" loop made of two perpendicular squares touching along one side makes an even more interesting test than we suggested before. One could try to verify that mean vorticity is directed along the normal to the minimal surface anywhere at this surface, not just at the edge. Here is the minimal Surface for Soccer Gates loop, created with \Mathematica\ package \cite{MinSurfaceX}(see Fig. \ref{fig:SoccerGates}).

 \section{Acknowledgements}
 I am indebted to Katepalli R. Sreenivasan, Kartik P. Iyer and Victor Yakhot for stimulating discussions. Arthur Migdal helped me with \Mathematica\ programming.
 
\bibliographystyle{unsrt}  


\end{document}